\documentclass[12pt]{article}

\usepackage{epsfig,graphics}

\newcommand{\p}{\partial}
\newcommand{\al}{\alpha}
\newcommand{\la}{\lambda}
\newcommand{\de}{\delta}
\newcommand{\ds}{\displaystyle}

\newcommand{\lf}{\left}
\newcommand{\rt}{\right}

\newcommand{\hgt}{\hat{\tau}}
\newcommand{\hnu}{\hat{\nu}}
\newcommand{\hA}{\hat{A}}
\newcommand{\tit}{\textit}
\newcommand{\tbf}{\textbf}

\newenvironment{reflist}{\begin{list}{}{\listparindent -.15in
  \leftmargin .5in} \item \ \vspace{-.35in} }{\end{list} }

\begin{document}
\title{Extrapolation of Threshold-Limited\\ Null Measurement Frequencies}
\author{O.~E.~Percus\\
Courant Institute/NYU \\
251 Mercer Street \\
New York, NY  10012 \\
\and
J.~K.~Percus \\
Courant Institute/NYU \\
251 Mercer Street \\
New York, NY  10012 \\
}
\date{}
\maketitle


\newpage

\begin{abstract}
The total measurable level of a pathogen is due to many sources, which produce
a variety of pulses, overlapping in time, that rise suddenly and then decay.

What is measured is the level of the total contribution of the sources at
a given time.  But since we are only capable of measuring the total level
above some threshold $x_0$, we would like to predict the distribution below
this level.

Our principal model assumption is that of the asymptotic exponential decay
of all pulses.  We show that this implies a power law distribution for the
frequencies of low amplitude observations.  As a consequence, there is a
simple extrapolation procedure for carrying the data to the region below
$x_0$.
\end{abstract}


\section{Introduction}

Acquiring sufficient data of sufficient accuracy is the standard problem in
the use of applicable mathematics.  Reliance upon null measurements---i.e.,
an answer of yes or no---is often an intelligent way of attending to the
latter desideratum, as in the familiar limiting dilution assays
[Lefkowitz and Waldman, 1979].  But the former frequently is controlled by
experimental inability, or perhaps excessive expense, in dealing with some
region of data.  If enough is surmised about the structure of the data,
such regions can be reduced by suitable extrapolation, but the implicit
assumption [for an elegant presentation, see Berman, 2006] of some sort of analytic structure runs the risk
of being too much of a mathematical band-aid unless it is justified by
a versatile underlying model.

In this note, we address a situation of some generality.  It is that in which
an organism, biological or mechanical, is continually subjected to transient
defects, e.g.\ pathogenic molecular species, internally or externally incited
but soon eliminated.  These inhibit its ability to effectively deal with
its environment.  We imagine that the net pathogen level $A$ is measurable at
occasional time intervals, but only if it exceeds some threshold $x_0$
(i.e.\ $A \ge x_0$).  A null measurement sequence would then give the relative
frequency $G(x_0)$ of measurements falling below the threshold $x_0$.
We would want e.g.\ to obtain from this the density function $\rho(A)$
of amplitudes of the pathogen aggregate level, $A$, with particular attention
to the unavailable low amplitudes.  The total pathogen load $A$ at a given
measurement would be expected to be the resultant of the current amplitudes
of each of the sources;  these sources may be imagined as time-displaced
versions of a discrete set of types, and this is the model that we will
study in detail.  The model was originally used in a somewhat different context,
that of the significance of ``blips'' in HIV viral level in patients undergoing
multi-drug therapy. [see~Percus et al, 2003]

What we can adjust in this scenario is the threshold level \textit{above}
$x_0$, and then observe the null frequency $G(x)$ for $x \ge x_0$. The
relationship between the intrinsic $\rho(A)$ and $G(x)$ is obvious
\begin{equation}\label{1.1}
G(x) =\int^x_0 \rho(A)\,dA,
\end{equation}
just the cummulative distribution of $A$.  Our task is now to obtain the form
of $\rho(A)$ from the model assumptions and use this e.g.\ to extrapolate
the available $G(x)$ for $x \ge x_0$ to values  $0<x< x_0$.

\section{The Underlying Model}

\begin{figure}[htbp]
\epsfig{figure=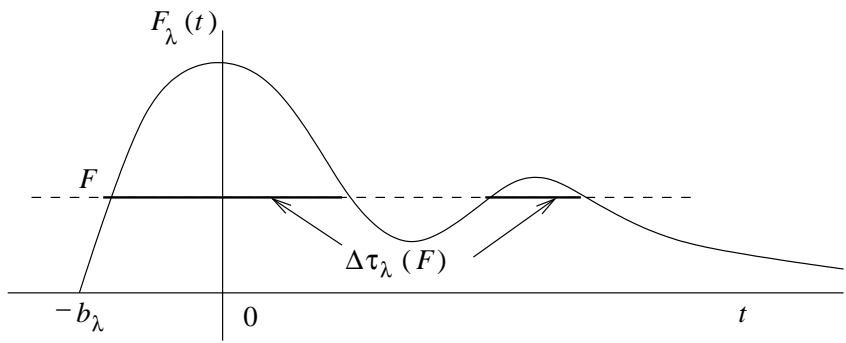}
\caption{Parameters of Typical Pulse Shape}
\end{figure}

We imagine that the arriving pulses are all translations in time of a basic set of
shapes indexed by $\la$
\begin{equation}\label{2.1}
F_\la(t), \; a<t<b
\end{equation}
These shapes are non-negative functions such that
$$
\int^b_a F_\la(t) \, dt\quad \hbox{ is  finite,}
$$
Now, place each of these functions, independently on the interval $(-T, T) (-T<a<b<T) \,\nu_\la$
times. To do this, let $\hgt$ be a random variable uniformly distributed on the interval
$(-T,T)$ and $\hnu_\la$ a Possion random variable with mean $2T q_\la$, i.e.
\begin{equation}\label{2.2}
h_\la(\nu_\la) \equiv P\{\hnu_\la =\nu_\la\} =
\frac{(2T q_\la)^{\nu_\la}}{\nu_\la!}
\;e^{-2T q_\la}.
\end{equation}
The location in time of $F_\la$ is determined, for example, so that its maximum
is at the origin (see Fig.~1). The time coordinate of the maximum point of the $i^{th}$
occurrence of $F_\la$ is  then denoted by $\hgt_{\la_i}$ $i=1, \dots, \, \hnu_\la$.

The equation of the $i^{th}$ occurrence of the curve $F_\la$ is then
$$
\hA_{\la_i}= F_\la \lf(t-\hgt_{\la_i}\rt).
$$
The total amplitude at any specified time, say $t=0$, is
\begin{equation}\label{2.3}
\hA = \sum_\la \sum^{\hnu_\la}_{j=1} F_\la\lf(-\hgt_{\la j}\rt).
\end{equation}
We would like to find the probability density of the random variable
$\hA$.

Let $\rho(A)$ be the probability density function of the random variable
$\hA$ i.e.
\begin{equation}\label{2.4}
\rho(A)=(\p/\p A) Pr \,\big(\hA \le A\big)
\end{equation}
We will assume a steady state distribution of ``pathogens'' in the course
of measurements.  This is a limitation of our approach: often the life-time
of the organism may be comparable to the ``decay'' of pathogen.
Then the system is translation-invariant in time, which is why we can
choose, without loss of generality, the observation time $t=0$,
as in (\ref{2.3}).

Let us construct the generating function for $\rho(A)$
\begin{equation}
w(\al) \equiv E\lf(e^{-\al \hA}\rt) \label{2.5}\\
\end{equation}
Then
\begin{equation}
w(\al)= \ds{\int^\infty_0 e^{-\al \,A} \;\rho(A)\,dA.}\label{2.6}
\end{equation}
We need
\begin{equation}\label{2.7}
E\lf(e^{-\al\, F_\la(-\hgt_{\la_j})}\rt) = \frac{1}{2T}
\int^T_{-T} e^{-\al \, F_\la(\tau)}\,d\tau,
\end{equation}
so that
\begin{equation}\label{2.8}
w(\al) = E \lf(\prod_\la \lf(\frac{1}{2T} \int^T_{-T} \, e^{-\al
\, F_\la(\tau)}\,d\tau\rt)^{\nu_\la} \rt).
\end{equation}
But from (\ref{2.2}), $E\lf(Y^{\hnu_\la}\rt) = e^{2T\,q_\la(Y-1)}$, and we see
at once that
\begin{equation}\label{2.9}
w(\al) = \prod_{\la} \exp \lf[ q_\la \int^T_{-T}
\lf(e^{-\al\,F_\la(\tau)}-1\rt) d\tau\rt],
\end{equation}
or letting $T\to \infty$,
\begin{equation}\label{2.10}
\int^\infty_0 e^{-\al \, A}\; \rho(A)\, dA = \exp \sum_\la
\lf[q_\la \int^\infty_{-\infty} \lf(e^{-\al \, F_\la (\tau)}-1\rt)d\tau\rt],
\end{equation}
which is our basic expression.

Eq.~(\ref{2.10}) can be expressed more concisely.
Define $\Delta\tau_\la(F)$ (see Fig.~1) as the total time that the ordinate
$F_\la(\tau) \ge F$ i.e.\ $\Delta\tau_\la(F) \equiv
\displaystyle{\int\theta \lf(F_\la(\tau)
-F\rt)d\tau}$ where $\theta(x) = \Bigg\{\begin{array}{ll}
0 &\mbox{if $x < 0$}\\[-.4cm]
1 &\mbox{if $x\ge 0$}
\end{array}$.

Also note that $\Delta \, \tau'_\la(F) \equiv \frac{d}{dF} \: \Delta
\tau_\la(F) =- \de \lf(F_\la (\tau)-F\rt)$ where $\delta (x)$ is the
Dirac $\delta$ function.  Then for any function $f$ we have
$$
\int f(F) \, \Delta \tau'_\la(F) \,dF=- \int f(F) \int \delta
\lf(F_\la (\tau)-F\rt) d\tau dF=-\int f\lf(F_\la(\tau)\rt)d\tau.
$$
It follows that
\begin{equation}\label{2.11}
\int^\infty_{-\infty} \lf(e^{-\al\,F_\la(\tau)}\,-1\rt) d\tau=
\int^\infty_0 \lf(1-e^{-\al \, F}\rt)\Delta \tau'_\la(F)\,dF,
\end{equation}
so that if
\begin{equation}\label{2.12}
\tau(F) \equiv \sum_\la q_\la \,\Delta \tau_\la(F),
\end{equation}
we have the simple equality
\begin{equation}\label{2.13}
\int^\infty_0 e^{-\al\,A} \rho(A)\, dA = \exp \int^\infty_0
\lf(1-e^{-\al\, F}\rt)\tau'(F)\,dF .
\end{equation}

\section{Rationale for Extrapolation}

Eq.~(\ref{2.13}) can of course be solved for $\rho(A)$ in nominal closed
form by applying the inverse Laplace transform.  But a less formal path
is to use (\ref{2.13}) to set up an equation that $\rho(A)$ satisfies.
For this purpose, take the logarithm of the equality (\ref{2.13}) and
apply the operation $-\p/ \p \al$ to both sides, yielding

\begin{equation}\label{3.1}
\begin{array}{lll}
\ds{\int^\infty_0 e^{-\al\,A}\;A\,\rho(A) \, dA} &=\ds{ \int^\infty_0
e^{-\al \, F} \lf(-F\tau'(F)\rt)dF \int^\infty_0
e^{-\al\,A}\,\rho(A)\,dA}\\
&=\ds{ \int^\infty_0 \int^\infty_0 Q(F) \;e^{-\al\, (F+A)}
\:\rho(A)\, dA\, dF}\\
&=\ds{ \int^\infty_0 \int^\infty_0 Q(F) \;e^{-\al A}
\:\rho(A-F)\, dF\, dA}\\
\hbox{where } Q(F) &\equiv -F\,\tau' (F)
\end{array}
\end{equation}
and we have used the fact that $\rho(A)=0$ for $A<0$.  Now the inverse
Laplace transform (loosely, take the coefficient of $e^{-\al\,A}$ on both sides)
establishes that
\begin{equation}\label{3.2}
A\,\rho(A) = \int^A_0 Q(F)\, \rho(A-F)\,dF.
\end{equation}

Our interest  is in the behavior  of $\rho(A)$, or $G(X)$, for
small values of $A$, or $X$; since $F\le A$ in (3.2), this corresponds
to small values of $F$.  Now the anticipated nature of the pulse profiles
comes into play.  A pulse form of type $\la$ will be initiated (see Fig.~1)
at some time $-b_\la$.  If it is thereafter determined by any standard
chemical kinetic sequence leading to its eventual disappearance, it will
asymptotically decay as $C_{\la} \;e^{-a_\la t}$ for some $a_\la$.
Hence the low amplitude
level $F$ duration will be given by
\begin{equation}\label{3.3}
\tau_\la (F) = -b_\la-\frac{1}{a_\la} \;\ell_n (F/C_\la).
\end{equation}
Consequently, we have for the total weighted duration
\begin{equation}\label{3.4}
\begin{array}{lll}
\tau(F) =&\ds{\sum_\la q_\la \lf(-b_\la + \frac{1}{a_\la}\;\ell n\, C_\la\rt)}\\
&\qquad \ds{-\lf(\sum_\la \frac{q_\la}{a_\la}\rt)\ell n \, F,}
\end{array}
\end{equation}
from which $Q(F)$ of (3.1) has the constant value
\begin{equation}\label{3.5}
Q(F)=Q\equiv \sum_\la q_\la/a_\la.
\end{equation}
Eq.~(\ref{3.2}), with $\rho(F) =0$ for $F<0$ then becomes
\begin{equation}\label{3.6}
A\, \rho(A) =Q \int^A_0 \rho(F)\,dF,
\end{equation}
or in terms of the null measurement cumulant $G(x)$ of (1.1),
$x\,G'(x) = Q\, G(x)$, with the solution
\begin{equation}\label{3.7}
G(x)=Cx^Q.
\end{equation}
We conclude that
\begin{equation}\label{3.8}
\ell n \, G(x) = \ell n \, C+Q\, \ell n \, x,
\end{equation}
so that a standard linear extrapolation of $\ell n\, G$ vs $\ell n \, x$
is valid at sufficiently small $x$

Let us take a hypothetical example.  It is that of chronic parasitic
infection of an organism, with continual birth of clusters of parasites,
each of which is quenched by the immune system.  There is a large
fluctuation in parasite load $A$, sampled sequentially in equivalent test
volumes, measurable if above the threshold $x_0$.  If the data is
acquired via null measurements of the load above virtual thresholds
$\{x \ge x_0\}$, we want to extrapolate the ensuing $G(x)$ to $x< x_0$.
Choose as typical population spike, (with origin at $\tau=0$ rather than
at $\max F_\la$---it makes no difference) the form
\begin{equation}\label{3.9}
F_\la (-\tau) = C_\la\, e^{-a_\la\, \tau} \lf(1-e^{-d_\la \, \tau}\rt),
\end{equation}
and for definiteness, $1 \le c_\la \le 5$, $1 \le a_\la \le 3$,
$1\le d_\la \le 5$ over a period $0\le \tau \le 10$, with parameters
distributed uniformly in their domains, and all $q_\la=1$.  Evaluating
$A$ of Eq.~(\ref{2.5}) for 1000 runs, the resulting $\ell n \,G(x)$ is plotted
against $\ell n\, x$ in Fig.~2.  The feasible linear extrapolation region
is indeed very large.

The conclusion (\ref{3.8}) is not without assumptions that have been pointed
out, but it appears to be a result of some generality, exemplifying the
assertion that extrapolation is a model-dependent procedure, and that
recognition of this fact has important operational significance.

\begin{figure}[htbp]
\epsfig{figure=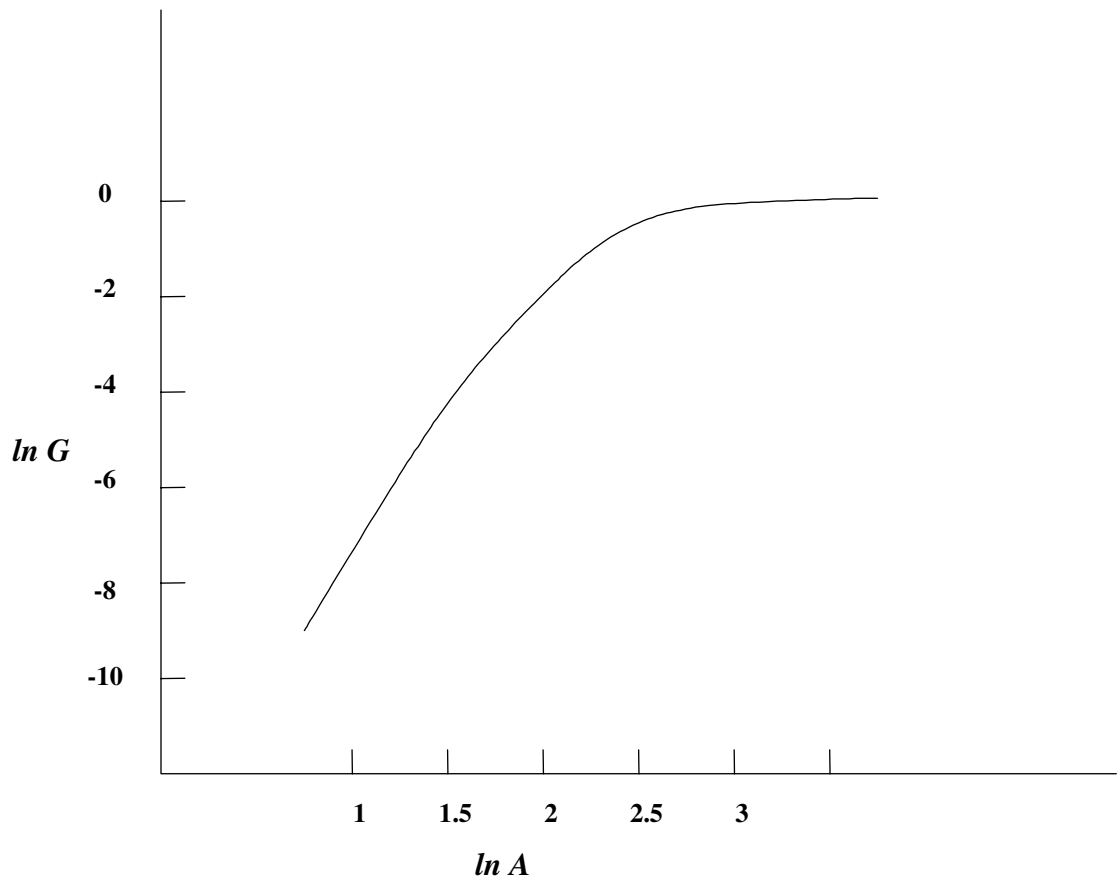}
\caption{Typical Dependence of $\ell n\; G$ on $\ell n\; A$}
\end{figure}

\newpage

\section*{References}

\begin{reflist}

Berman, S, 2006.  Legendre Polynomial Kernel Estimation,
\tit{Comm.\ Pure and Appl.\ Math.} \tbf{60}, 1238.

Lefkowitz, I and Waldman, H, 1979.  Limiting Dilution Analysis
of Cells in the Immune System, \tit{Cambridge Press}

Percus, Percus, Markowitz, Ho, di Mascio, and Perelson, 2003.
The distribution of viral blips observed in HIV-1, \tit{Bull.\ Math.\ Bio.}
\tbf{65}, 263--277.

\end{reflist}

\end{document}